\documentclass[reprint, superscriptaddress, aps]{revtex4-2}

\usepackage{newtxtext}
\usepackage{newtxmath}
\usepackage[T1]{fontenc}

\usepackage{amsmath}
\usepackage{nicefrac}
\usepackage{braket}
\usepackage{graphicx} 
\usepackage{placeins}
\usepackage[hidelinks, colorlinks = false]{hyperref}

\newcommand{\abs}[1]{\left| #1 \right |}

\def\figref#1{Fig.~\ref{#1}}
\def\tabref#1{Tab.~\ref{#1}}

\def\pihalf{\nicefrac{\pi}{2}}

\def\emath{e}
\def\QED{\hfill$\blacksquare$}
\def\re#1{\mathop{\mathrm{Re}}(#1)}
\def\im#1{\mathop{\mathrm{Im}}(#1)}

\begin{document}

\title{Generalization of catability for parity-less cat states}

\author{Michal Matul\'ik}
\email{matulik@optics.upol.cz}

\author{Jan Provazn\'ik}
\email{provaznik@optics.upol.cz}
\author{Petr Marek}

\email{marek@optics.upol.cz}
\affiliation{Department of Optics, Palack\'y University, 17. listopadu 1192/12, 771 46 Olomouc, Czech Republic}

\author{\v{S}imon Br\"auer}
\email{brauer@optics.upol.cz}
\affiliation{Niels Bohr Institute, University of Copenhagen, Jagtvej 155 A, DK-2200 Copenhagen, Denmark}

\begin{abstract}\noindent
We extend the nullifier-based certification framework of catability from parity-defined coherent cat states to parity-less Kerr cat states. Since Kerr cats cannot be distinguished by parity alone, we replace the parity component of the original nullifier with a displaced-parity operator that captures their characteristic interference structure. The resulting generalized catability remains directly observable and can be evaluated from a finite number of photon-number measurements without full state tomography. Numerical benchmarks show that the method faithfully certifies Kerr-cat features, produces accurate approximations of ideal Kerr cat states, and is more resilient to optical loss than fidelity-based certification. We also identify an exact anti-linear nullifier based on complex conjugation, providing an ideal algebraic description of Kerr cat states. Our results broaden the scope of catability and provide an experimentally practical approach to the characterization of Kerr cat states.
\end{abstract}

\maketitle

\section{Introduction}
\label{sec:intro}


Coherent cat states~\cite{schrdinger,dodonov,yurke} are known as versatile resources on various quantum platforms with applications in quantum information encoding schemes~\cite{jeong,ralph,lund2008,su,hastrup2022,lee2024}, quantum metrology~\cite{duivenvoorden2017,pan2025}, and state-engineering protocols~\cite{lund,sychev2017,weigand2018,konno,hastrup,banic}. Cat states of assorted quality have been experimentally prepared in traveling optical fields~\cite{ourjoumtsev2006,neergaardnielsen2006,jezek,neergaardnielsen2006,endo2023,endo2025}, Kerr-nonlinear resonators~\cite{puri2017,grimm2020, he2023,ding2025}, trapped atoms~\cite{Omran2019Aug} and ions~\cite{Monroe1996May,Kienzler2016Apr}, in optomechanical~\cite{Shomroni2020Mar,Hauer2023May,Bild2023Apr} and hybrid light-matter systems~\cite{Hacker2019Feb}.

Reliable characterization of the prepared cat states remains a challenging task. Quantum non-Gaussianity~\cite{filip2011,walschaers2021,lachman2022}, one of the defining features of coherent-state superpositions, can be identified with diverse indicators, including Wigner-function negativity, measures based on distance and entropy, moment criteria, and other operational witnesses~\cite{Lee2011May,fiurasek2013,park2015,Kwon2017Apr}. These approaches are complementary as they certify different aspects of the state and generally do not establish a single hierarchy of non-Gaussian resources. 

While fidelity with an ideal target state is often used because it gives a direct comparison between the experimental state and its theoretical reference~\cite{sychev2017,he2023,bild2023}, it does not reveal which physical features are responsible for the reduced overlap displayed by imperfect states and its evaluation usually requires full-state reconstruction~\cite{neergaardnielsen2006,he2023,endo2023,endo2025,genoni2013}.

Alternatively, coherent cat states can be characterized with catability~\cite{catability}, which identifies their intrinsic interference structure. This quantifier is based on a pair of directly observable nullifiers, one identifying the coherent contributions, and one that ensures the state exhibits the correct parity. Its expectation value, normalized by the best value attainable by Gaussian states, defines nonlinear squeezing~\cite{marek2024,kala2025,brauer2025}, which serves as a witness of cat-like non-Gaussianity and connects state certification to measurable quantities rather than to full-state reconstruction~\cite{endo2023,sonoyama2023,endo2021}. 

Catability in its original form~\cite{catability} was restricted specifically to two-headed cat states with well-distinguishable parities, which are only a special case within the set of balanced superpositions of coherent states with complex coefficients. The set also contains the family of parity-less Kerr cats ${\ket{\alpha} \pm i \ket{-\alpha}}$ that cannot be differentiated from each other by parity alone. 


The present work extends the concept of catability to include Kerr cat states. We address its limitations by considering a displaced parity operator instead in the second nullifier. However, the ground states of the modified operator are not exact Kerr cat states, which we solve by introducing an anti-linear nullifier for the ideal Kerr cat state, which gives a compact algebraic characterization of the perfect case.

The paper is organized into five major sections, starting with the anti-linear nullifier of Kerr cats in the \hyperref[sec:cats_and_nullifiers]{second section}, followed by the introduction of a physically meaningful displaced parity operator in the \hyperref[sec:approximate_kerr_cats]{third section} and an analysis of its ground states within the \hyperref[sec:numerical_benchmark]{fourth section}. The generalized catability is defined in the \hyperref[sec:catability]{fifth section} and its performance is compared with the standard approach based on fidelity. A practical decomposition requiring only a limited number of measurements is discussed in the \hyperref[sec:practical]{final section} of the manuscript.

\section{Kerr cat states and their nullifiers}
\label{sec:cats_and_nullifiers} 

\begin{figure}[t]
    \centering
    \includegraphics[width=0.42\textwidth]{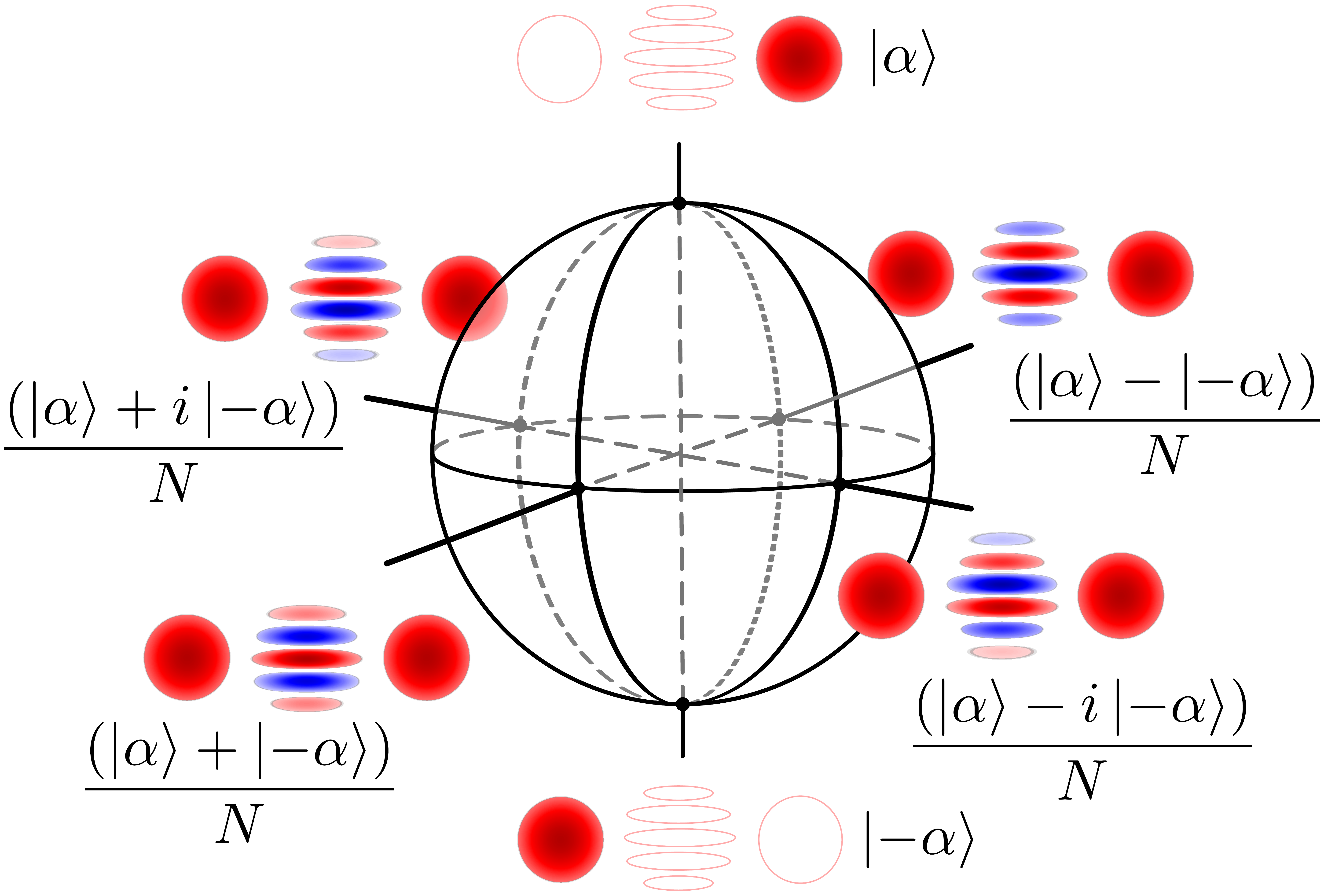}
    \caption{
        Illustration of a Bloch sphere with logical qubits formed by superpositions of coherent states.
        Its poles correspond to the individual coherent components~$\ket{\alpha}$ and~$\ket{-\alpha}$, while the equal-weight coherent-state superpositions lie on the equator. In the notation of~\eqref{eq:general_cat}, the conventional even and odd parity cats occupy opposite equatorial points, while the Kerr cat states occupy the orthogonal equatorial points. The Wigner functions surrounding the sphere illustrate the corresponding base states.
    }
    \label{fig:bloch_sphere}
\end{figure}

We consider the family of equatorial cat states, 
\begin{equation}\label{eq:general_cat}
    \ket{C_{\theta}(\alpha)} =  \frac{1}{\sqrt{M}}\big (\ket{\alpha} + \emath^{\imath \theta} \ket{-\alpha} \big ),
\end{equation}
where $\sqrt M$ denotes the normalization factor
\begin{equation}\label{eq:norm}
    M = 2\big(1 + \cos{(\theta})\emath^{-2 \abs{\alpha}^2} \big).
\end{equation}
The term \emph{equatorial} is inspired by the standard qubit representation, illustrated in~\figref{fig:bloch_sphere}, and motivated by the possibility of encoding quantum information in superpositions of cat states~\cite{puri2017,grimm2020, he2023,ding2025}. The notation~\eqref{eq:general_cat} fixes the relative phase between the two coherent states and provides a common basis for discussing both conventional cat states with well-defined even and odd parities, and parity-less Kerr cats.

The parity of conventional cat states, with the relative phases~${\theta\in\{0,\pi\}}$, reflects that the states contain only even or odd photonic contributions. Conversely, the parity-less Kerr cats, obtained for ${\theta=\pm\pihalf}$, incorporate all contributions
\begin{align}\label{eq:kerr_cat_1}
   \ket{C_{+ {\pihalf}} (\alpha)} & = (1+\imath)\ket{0} + (1-\imath )\alpha \ket{1} \nonumber\\
   &+ \frac{\alpha^2}{\sqrt{2}} (1+\imath) \ket{2}+\cdots,\\ \label{eq:kerr_cat_2}
   \ket{C_{- {\pihalf}} (\alpha)} & = (1-\imath)\ket{0} + (1+\imath) \alpha \ket{1} \nonumber\\
   &+ \frac{\alpha^2}{\sqrt{2}} (1-\imath) \ket{2}+\cdots
\end{align}
and consequently do not exhibit a specific parity. Please note the complex conjugation between \eqref{eq:kerr_cat_1} and \eqref{eq:kerr_cat_2}. 

The original catability quantifier~\cite{catability} is based on the nullifier 
\begin{equation}\label{eq:operator}
    \hat{O} (\alpha, \gamma, s) 
    \equiv \hat{O}_{C}(\alpha) + \hat{O}_{P}(\gamma, s)
\end{equation}
comprising the amplitude $\hat{O}_{C}$ and parity $\hat{O}_{P}$, where
\begin{equation} \label{eq:operator_alpha}
    \hat{O}_{C} (\alpha) 
    \equiv \big ({\hat{a}^{\dagger 2}}  - {\alpha^{*2}} \big )\big (\hat{a}^2  - \alpha^2 \big )
\end{equation}
has a nullspace spanned by coherent states~$\ket{\alpha}$ and~$\ket{-\alpha}$. The symbols~$\hat{a}$ and~$\hat{a}^{\dagger}$ are the annihilation and creation operators of the harmonic oscillator \cite{nielsen}. The parity term is defined as
\begin{equation}\label{eq:operator_sub}
     \hat{O}_{P}(\gamma,s) \equiv \gamma\big ( 1 -s \hat{\Pi} \big ),
\end{equation}
where ${s = \pm 1}$ represents the value of~${\emath^{\imath \theta}}$ within the superposition~\eqref{eq:general_cat}. The sign of $s$ determines whether the operator becomes a nullifier for even~(${+1}$) or odd~(${-1}$) cat states. The parity operator~$\hat{\Pi}$ itself, which can be defined as 
\begin{equation}\label{eq:operator_parity}
    \hat{\Pi} = \emath^{-\imath \pi \hat{n}},
\end{equation}
where ${\hat{n} = \hat{a}^{\dagger} \hat{a}}$ represents the photon number operator, realizes a $\pi$-rotation in the phase space. The expectation value, taken with respect to some state $\hat{\rho}$, corresponds to the central point 
\begin{equation}
    \langle \hat{\Pi} \rangle = \pi W_{\hat{\rho}}(0),
\end{equation}
of its Wigner function $W_{\hat{\rho}}(\xi)$~\cite{royer}.


The main difficulty in adapting the relation~\eqref{eq:operator} for Kerr cats lies in their structure. Both states, \eqref{eq:kerr_cat_1} and~\eqref{eq:kerr_cat_2}, have identical expectation values of the parity operator, which cannot be used to differentiate between them. 

We can solve this problem by replacing the inadequate parity operator~$\hat{\Pi}$ in the nullifier~$\hat{O}_{P}$ with the anti-linear operator~$\hat{K}$ of complex conjugation \cite{wigner,cejnar2015,sakurai,ballentine}, defined by its action 
\begin{equation}\label{eq:KK}
    \hat{K} c = c^* \hat{K}, \; \; \; c \in \mathbb{C}.
\end{equation}
Unlike the parity operator, it has no Wigner or Fock representation. It is not directly measurable either. It can only be used as an ideal reference. Its action can be approximated with measurable displaced parity operators \cite{royer,bishop}.


The modified ${\hat{O}_{P} (\gamma, s) = \gamma (\imath -s \hat{K})}$ works only for Kerr cats with imaginary amplitudes. See Appendix~\ref{sec:app_a} for details.


\section{Displaced parity operator and approximate Kerr cat states}
\label{sec:approximate_kerr_cats}

Wigner functions of the parity-less Kerr cat states resemble those of the conventional cats. The major difference, as can be seen in~\figref{fig:bloch_sphere}, is that the interference fringes are shifted. Their phase space symmetries can be captured by the operator of complex conjugation~\eqref{eq:KK}, but this is impractical as the operator can not be measured directly. The fringes can be also targeted using a displaced parity operator, 
\begin{equation} \label{eq:displaced_parity_operator}
    \hat{\Pi}^{(\beta)} 
    \equiv \hat{D}(\beta) \hat{\Pi} \hat{D}^{\dagger}(\beta) 
    = \emath^{\imath \pi (\hat{a}^{\dagger} - \beta^*)(\hat{a} - \beta) },
\end{equation}
where ${\hat{D}(\beta)}$ represents the displacement operator
\begin{equation}\label{eq:displacement_operator}
    \hat{D}(\beta) 
    = \emath^{\beta \hat{a}^{\dagger} - \beta^* \hat{a}}
\end{equation}
and $\beta \in \mathbb{C}$ gives the displacement amplitude. Unlike \eqref{eq:KK}, the displaced parity operator is measurable. Its expectation value
\begin{equation}\label{eq:displaced_parity}
    \langle \hat{\Pi}^{(\beta)} \rangle  
    = \pi W_{\hat{\rho}} (\beta).
\end{equation} 
corresponds to the value of the Wigner function at some point~$\beta$ of the phase space~\cite{royer,bishop}. The modified operator $\hat{O}_{P}$ reads
\begin{equation}\label{eq:approx_parity_operator}
    \hat{O}_{P} (\beta,\gamma,s) 
    = \gamma \bigg(1 - s\hat{\Pi}^{(\beta)}  \bigg ),
\end{equation}
where ${s = \pm 1}$. The appropriate displacement amplitude~$\beta$ for a general equatorial state~\eqref{eq:general_cat} can be determined by solving
\begin{equation}\label{eq:extremum}
    \mathop{\text{minimize}}
    \limits_{\beta} f(\alpha, \beta, \theta)
\end{equation}
where ${f(\alpha, \beta, \theta)}$ is the expectation value~\eqref{eq:displaced_parity} calculated as
\begin{equation}\label{eq:expectation_of_parity}
    \begin{aligned}
        f(\alpha, \beta, \theta) 
        & = \braket{
            C_{\theta}(\alpha) | 
            \hat{\Pi}^{(\beta)} |
            C_{\theta}(\alpha) 
        } \\
        & = \frac{g(\alpha, \beta, \theta) + g(\alpha, - \beta, - \theta)}{M} \\
        g (\alpha, \beta, \theta) 
        & = \emath^{-2 \abs {\alpha-\beta}^2} 
            + e^{i \theta} \emath^{
                -2 \abs \beta^2 + 2 \beta \alpha^* - 2 \beta^* \alpha 
            }
    \end{aligned}
\end{equation}
where~$M$ is the square of the normalization factor~\eqref{eq:norm}.
The relation at hand can be simplified without loss of generality. We can rotate the coherent states to ensure~${\re{\alpha} = 0}$ and further assume~${\im{\beta} = 0}$ in the displacement, resulting in
\begin{equation}\label{eq:general_case_expecation_value}
    f (\alpha, \beta, \theta) 
    = 2 M^{-1} \emath^{-2 \abs \beta^2} 
        \big(  
            \emath^{-2 \abs \alpha ^2} +
            \cos{(\theta  -4 \beta \alpha_R)}  
        \big),  
\end{equation}
where we substituted ${\im{\alpha} = \imath \alpha_{R}}$ with ${\alpha_{R} \in \mathbb{R}}$. 
In the case of the Kerr cat states, the relation~\eqref{eq:general_case_expecation_value} further simplifies into
\begin{equation}\label{eq:special_caseB_expectation_value}
    \begin{aligned}
        f\left(\alpha, \beta, \pm \pihalf\right) 
        & = \emath^{-2 \abs \beta^2 } \big( 
                \emath^{-2\abs\alpha^2} \pm \sin{(4\alpha_R\beta) }
            \big ) . 
    \end{aligned}
\end{equation}
We are unable to find the optimal value of~$\beta$ analytically because~\eqref{eq:special_caseB_expectation_value} is transcendental. Its value can only be determined numerically.
The optimal displacement for the conventional cat states is trivial with ${\beta = 0}$. In the general case of equatorial states, that is, when ${\theta \neq \{0,\pi,\pm\pihalf \}}$, the expectation value is given by~\eqref{eq:expectation_of_parity} and has to be solved for given~$\theta$ by any means necessary.

Once the value of $\beta$ is known, it can be used in~\eqref{eq:approx_parity_operator} and the ground state of the modified catability operator 
\begin{equation}\label{eq:total_approx_operator}
    \begin{aligned}
        \hat{O} (\alpha, \beta, \gamma,s) 
        & = \hat{O}_{C} (\alpha) + \hat{O}_{P} (\beta,\gamma,s) \\
        & = \big ({\hat{a}^{\dagger 2}}  - {\alpha^{*2}} \big )\big (\hat{a}^2  - \alpha^2 \big ) + \gamma \big(1 -s \hat{\Pi}^{(\beta)}  \big )
    \end{aligned}
\end{equation}
can be found numerically. For~${\theta = \pm \pihalf}$ the ground states approximate the respective Kerr cat states, satisfying 
\begin{equation}\label{eq:approx_kerr_ket}
    \bigg(\hat{O} (\alpha,\beta,\gamma,s) -\lambda_{min} \bigg)
     \ket{C^{apx}_{\pm \pihalf}(\alpha,\beta,\gamma,s)} = 0,
\end{equation}
where ${\ket{C^{apx}_ {{\pihalf}}(\alpha,\beta,\gamma,s)}}$ represents the approximate Kerr cat state associated with the minimal eigenvalue~$\lambda_{min}$. The nullifier is only approximate and, in general, ${\lambda_{\min} \neq 0}$. The sign~$s$ controls which central fringe becomes displaced, whether the negative~(${s = -1}$) or the positive~(${s = 1}$) one. The optimal displacement amplitude~$\beta$ depends on~$\alpha$. Its sign distinguishes between the two Kerr cat states with the convention
\begin{align}
    \label{eq:positive_approx}
    s &= + 1 \begin{cases}
        \beta < 0 \rightarrow  \ket{C^{apx}_{-\pihalf}},\\
        \beta > 0 \rightarrow  \ket{C^{apx}_{ \pihalf}}.
    \end{cases} \\
    \label{eq:negative_approx}
    s&=-1 \begin{cases}
        \beta < 0 \rightarrow  \ket{C^{apx}_{ \pihalf}}, \\
        \beta > 0 \rightarrow  \ket{C^{apx}_{-\pihalf}},
    \end{cases}
\end{align}
An interesting limiting case is obtained for~${\beta \equiv \alpha}$ which displaces the parity operator onto one of the coherent peaks. The operator~\eqref{eq:total_approx_operator} is then nullified by the constituent coherent states
\begin{equation}
    \bra{\pm \alpha} \hat{O}(\alpha,\alpha,\gamma,\pm1) \ket{\pm \alpha} = 0. 
\end{equation}
In conclusion, the modified operator~\eqref{eq:total_approx_operator} approximately covers all poles of the cat-qubit Bloch sphere.

\section{Benchmarking the approximate nullifier}
\label{sec:numerical_benchmark}

\begin{figure}[t]
    \includegraphics[width = \columnwidth]{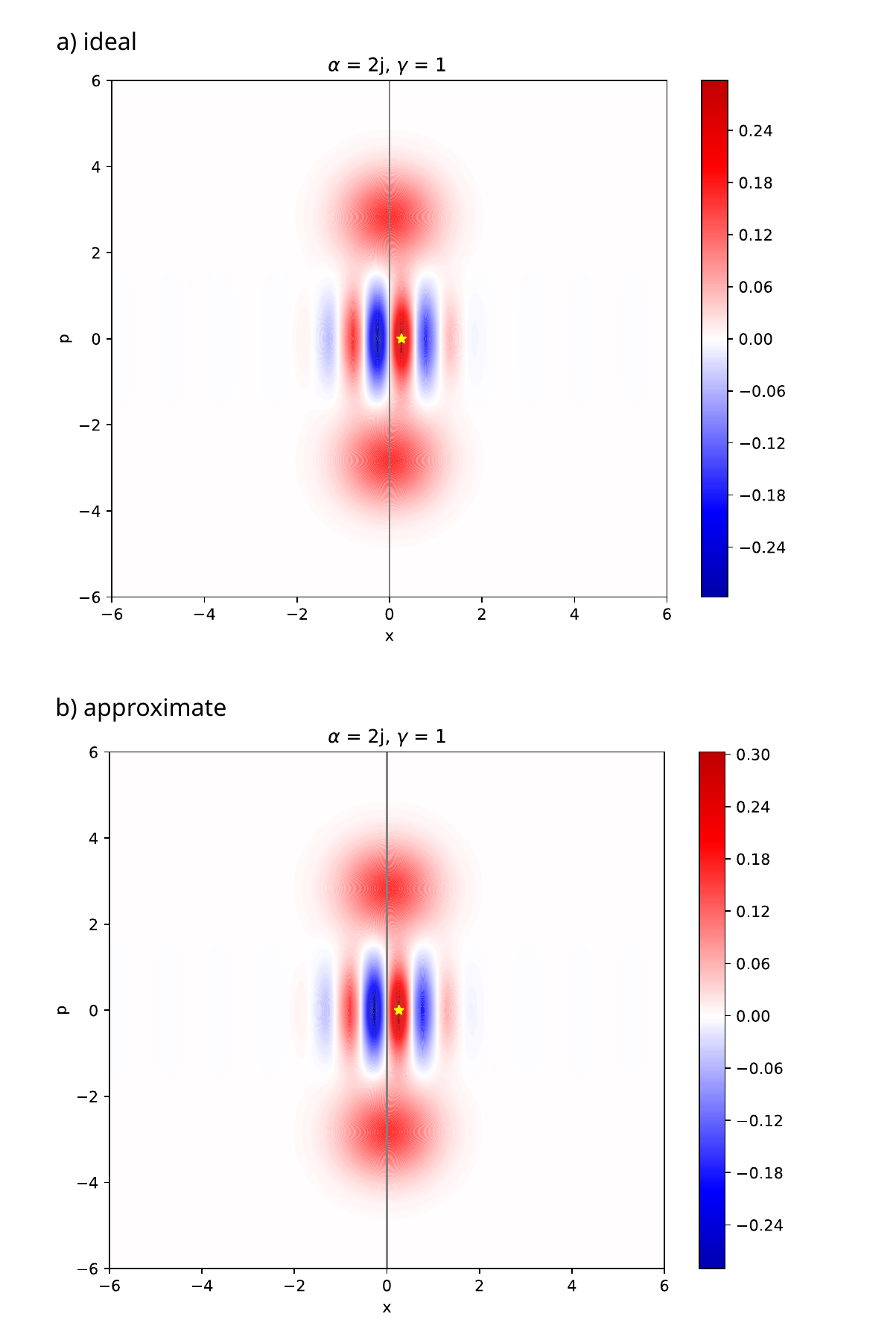} 
    \caption{
        Comparison of the ideal Kerr cat state~(a) and its approximation~(b) obtained as the ground state~\eqref{eq:approx_kerr_ket} for~${\alpha = 2i,\,\beta \approx \pm 0.185}$ and~${\gamma = 1,\, s = 1}$. The yellow star denotes the value of ${\beta \sqrt 2}$. Both states were constructed on a truncated Fock basis with dimension $90$ and their Wigner functions evaluated on a $473 \times 473$ phase space grid.
    } 
    \label{fig:wigner}
\end{figure}

The behavior of the approximate nullifier \eqref{eq:total_approx_operator} can be determined with numerical benchmarks. 
We begin by considering the differences between the ideal and approximate Kerr cat states with~${\alpha = 2i,\,\theta = \pihalf}$, the latter obtained as the ground state~\eqref{eq:approx_kerr_ket} with~${\beta\approx0.185}$. Even though their Wigner functions~${W(\xi)}$, illustrated in~\figref{fig:wigner}, appear to be visually indistinguishable from each other, there are slight differences, which are most pronounced in the center of the phase space and in the extremal points of their fringes. 

The quantitative differences between the ideal state and its approximations, obtained for different values of~$\gamma$, are summarized in~\tabref{tab:1}, which lists the central value~${W(0)}$ of the Wigner function, its values at the extremal points of the fringes and their respective positions. These values were determined from the discretized Wigner functions evaluated on discrete points of the phase space.
The subtle differences observed for the approximate states manifest because the operator~\eqref{eq:total_approx_operator} is only an approximate nullifier for Kerr cats and the ground states have non-zero eigenvalues $\lambda_{\min} \neq 0$. 

\begin{table*}[t]
    \caption{
        Summary of the quantitative differences between the ideal and approximate Kerr cat states with~${\alpha = 2i,\,\theta = \pihalf}$, the latter obtained as the ground state~\eqref{eq:approx_kerr_ket} for different values~${\gamma\in\{0.1, 1.0, 5.0\}}$ and~${\beta\approx0.185}$.
        The table lists the central value~$W(0)$ of the Wigner function and its values at the extremal points of the first positive~${W (x_{+})}$ and negative~${W (x_{-})}$ fringes, along with their respective positions $x_{+}$ and $x_{-}$.
    }
    \label{tab:1}%
    \begin{ruledtabular}
        \begin{tabular}{lrrrrr}
             &
            $W (x_{+})$ &
            $W (x_{-})$ &
            $x_{+}$ &
            $x_{-}$ &
            $W(0)$ \\
            \colrule
    
            Ideal &  0.295866 & -0.295666 &   0.254237 &  -  0.254237 & 0.000107 \\
            Approximate state ($\gamma = 0.1$) & 0.298724 & -0.292389  & 0.254237 &  -0.286017 & 0.029302\\
            Approximate state ($\gamma = 1.0$) & 0.302463 & -0.288056 &   0.254237 &  -0.286017 & 0.029282\\
            Approximate state ($\gamma = 5.0$) & 0.310717 & -0.274302 & 0.254237 &  -0.286017 & 0.029049\\  
        
        \end{tabular}
    \end{ruledtabular}
\end{table*}

\begin{figure}[t]
    \includegraphics[width = \columnwidth]{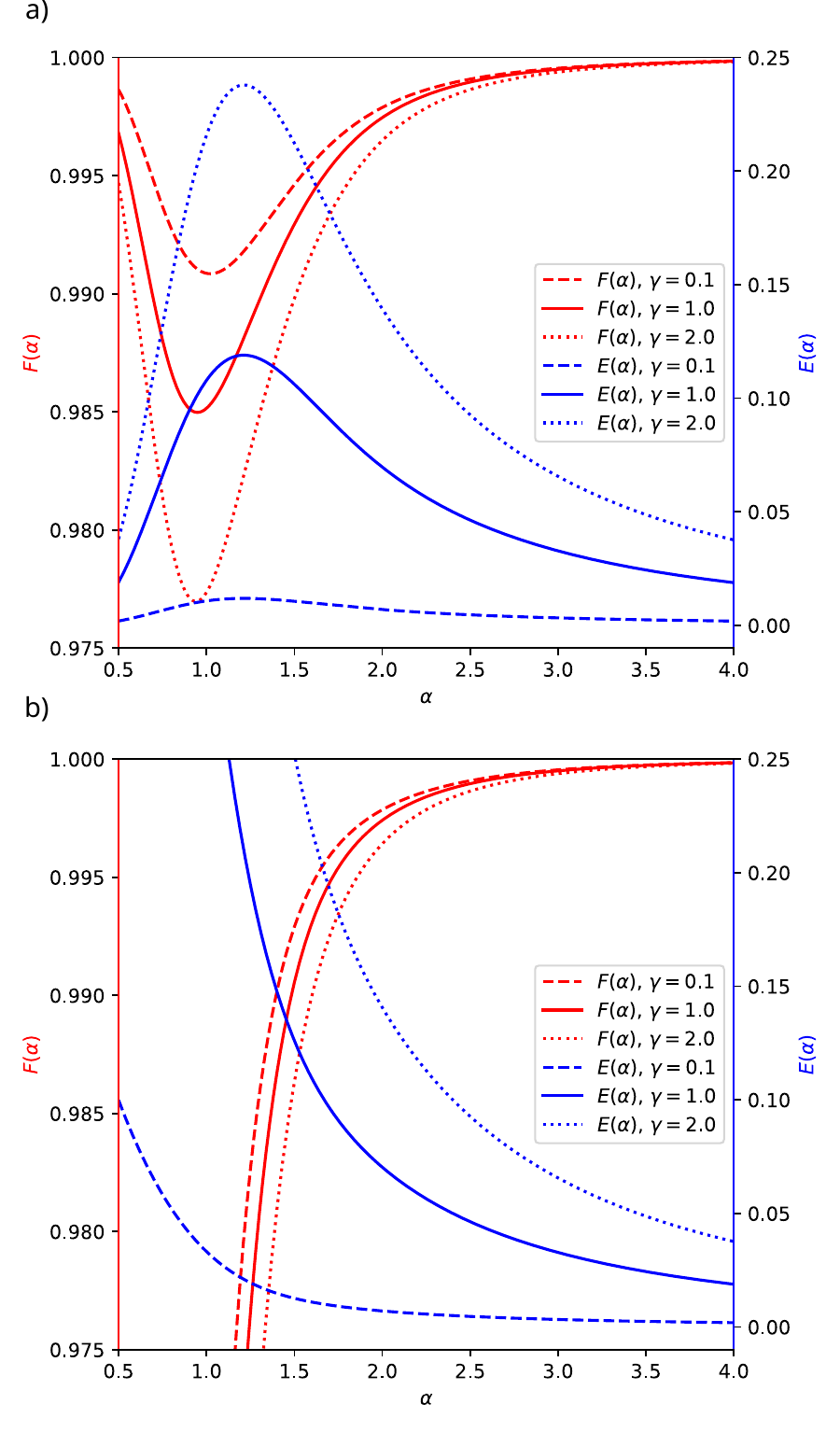} 
    \caption{
        The plots show the quantifiers \eqref{eq:fidelity} and \eqref{eq:expectation} as functions of the amplitude~$\alpha$ for displaced positive~(a) and negative~(b) fringes, as described in~\eqref{eq:positive_approx} and~\eqref{eq:negative_approx}. Fidelity, with vertical axis on the left, is shown in red, whereas blue portrays the expectation value of the nullifier with axis on the right. Dotted, solid, and dashed curves correspond to individual choices of ${\gamma \in \{ 0.1, 1.0, 2.0 \}}$. 
    } 
    \label{fig:comparison}  
\end{figure}

The quality of the approximation can be further assessed with a pair of complementary quantifiers, by using the fidelity of the approximate states with their ideal counterparts, and by computing the expectation values of the approximate nullifier with respect to ideal Kerr cat states. 
The two quantifiers,
\begin{align}
    \label{eq:fidelity}
    F (\alpha) 
    & \equiv \abs{ 
        \Braket{
              C_{\pihalf} (\alpha) 
            | C^{apx}_{\pihalf} (\alpha, \beta, \gamma, s)
        }
    }^2 \\
    \label{eq:expectation}
    E (\alpha)
    & \equiv \Braket{
          C_{\pihalf} (\alpha) 
        | \hat{O}(\alpha,\beta,\gamma,s) 
        | {C_{\pihalf}(\alpha)} 
    }
\end{align}
are evaluated in~\figref{fig:comparison} for a range of amplitudes~$\alpha$ and an assortment of~$\gamma$ factors. The optimal displacement~$\beta$ is determined for each amplitude. 
The plots show the quantifiers for approximate states, with the positive~(a) and negative~(b) central fringes displaced, corresponding to the relations~\eqref{eq:positive_approx} and~\eqref{eq:negative_approx}.
Fidelity~$F(\alpha)$ is depicted with red lines, whereas the blue lines show the expectation value~$E(\alpha)$. 

While the curves differ with~$\gamma$ within the limits of the plots, they converge in the limit of large~${\alpha \to \infty}$, where both quantifiers satisfy ~${F(\alpha) \to 1}$ and~${E(\alpha) \to 0}$ for all~$\gamma$. The approximation works well for sufficiently large amplitudes in both cases. The regime of small amplitudes offers a different interpretation. As~${\alpha \to 0}$, one has~${\beta \to 0}$ and, consequently,
\begin{equation}\label{eq:alpha_zero}
    \hat{O}(0, 0, \gamma,s) 
    = \hat{n}(\hat{n}-1) + \gamma \big(1 -s \hat{\Pi}^{}  \big),
\end{equation}
which has ground states~$\ket{0}$ (for~${s= 1}$) and~$\ket{1}$ (for~${s=-1}$) with zero eigenvalues~${\lambda_{\min} \equiv 0}$. The dominant component in ideal Kerr states~\eqref{eq:kerr_cat_1} is the vacuum state~$\ket{0}$, which is why we observe~${F(\alpha) \to 1}$ and~${E(\alpha) \to 0}$ in the first plot.

\section{Catability} 
\label{sec:catability}

\begin{figure}[t]
    \includegraphics[width = \columnwidth]{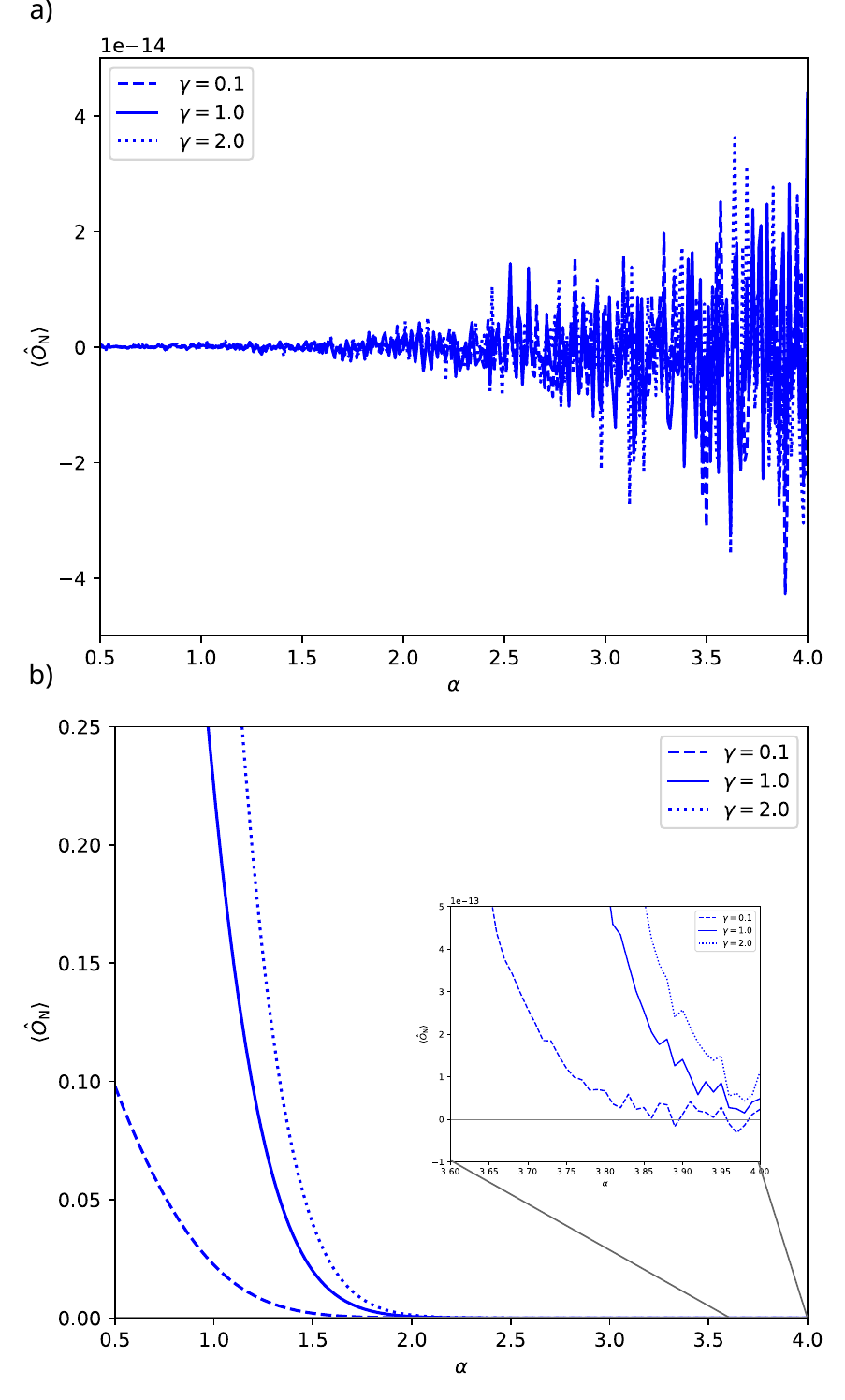} 
    \caption{ 
        The expectation value~\eqref{eq:expecation_normalized} of the improved nullifier~\eqref{eq:catability_operator} obtained for ideal Kerr cat states with varying amplitude~$\alpha$ and with their (a)~positive and (b)~negative fringes displaced. The dotted, solid, and dashed curves represent distinct~${\gamma\in \{ 0.1, 1.0, 2.0 \}}$ choices.
    } 
    \label{fig:expecation_norm}
\end{figure}

The original nullifier~\eqref{eq:operator} for the conventional cat states was based on the plain parity operator~\eqref{eq:operator_sub} and took advantage of the fact that the conventional cats satisfy~${\pi W(0) \equiv \pm 1}$, based on their parity, thus rendering the expectation value
\begin{equation}
    \langle 1 - s \hat{\Pi} \rangle
    \equiv 1 - s \pi W(0)
    \equiv 0
    .
\end{equation}
While the displaced parity operator~\eqref{eq:displaced_parity}, employed in the modified nullifier~\eqref{eq:total_approx_operator}, captures the essence of the approximate Kerr cat states quite well, it misses the mark for their ideal versions, since their expectation value
\begin{equation}
    \langle 1 - s \hat{\Pi}^{(\beta)} \rangle
    \equiv 1 - s \pi W(\beta)
    \neq 0,
\end{equation}
because~${\pi W(\beta) \neq \pm 1}$ for general values of~$\beta$. This can be resolved by using the actual value, obtained for the ideal states,
\begin{equation}
    \label{eq:normalization}    
    N^{(\beta)} 
    = \pi \abs{ W(\beta) }= \abs{
        \braket{
            C_{\pihalf}(\alpha)
            | \hat{\Pi}^{(\beta)}
            | C_{\pihalf}(\alpha)
        }
    },
\end{equation}
instead of the constant threshold in the generalized nullifier
\begin{equation}
    \label{eq:catability_operator}
    \hat{O}_{N} (\alpha,\beta,\gamma,s)
    = \hat{O}_{C} (\alpha) + \gamma \bigg(N^{(\beta)} - s\hat{\Pi}^{(\beta)}  \bigg ).
\end{equation}
We demonstrate the improvement in its behavior for the ideal Kerr cats in~\figref{fig:expecation_norm}, where we present the expectation value,
\begin{equation}\label{eq:expecation_normalized}
    \langle \hat{O}_{N} \rangle
    = \braket{
        C_{\pihalf}(\alpha)
        | \hat{O}_{N} (\alpha, \beta, \gamma, s)
        | C_{\pihalf}(\alpha)
    },
\end{equation}
obtained for the ideal states with their positive~(a) and negative~(b) fringes displaced. The former shows substantial improvement, with the expectation values effectively approaching zero, up to the extent of numerical precision. The latter case, with displaced negative fringe, shows similar improvement for larger amplitudes~$\alpha$. The expectation value~\eqref{eq:expecation_normalized} still diverges for smaller amplitudes, as was discussed previously for~\eqref{eq:alpha_zero}.

The generalized catability of a quantum state $\hat{\rho}$ is defined as
\begin{equation}\label{eq:catability_metric}
    \xi (\alpha) 
    \equiv 
    \min_{\gamma} 
    \frac{
        \text{Tr} \big(
            \hat{\rho}\hat{O}_{N}  (\alpha,\beta,\gamma,s) 
        \big)
    }{
        \min\limits_{\ket{\psi} \in \mathscr{H}} 
        \braket{\psi | \hat{O}_{N} (\alpha,\beta,\gamma,s) | \psi}
    },
\end{equation}
where the denominator serves as a normalization with respect to the class of all Gaussian states~$\mathscr{H}$. This makes the quantifier~\eqref{eq:catability_metric} a witness of non-Gaussianity as~${\xi (\alpha) <1 }$ necessarily indicates the state~$\hat{\rho}$ can not be Gaussian.
Catability~\eqref{eq:catability_metric} allows us to determine the quality of experimental cat states with known amplitudes~$\alpha$. We can also define its global version,
\begin{equation}\label{eq:global_catability}
    \xi = \min_{\alpha} \xi(\alpha),
\end{equation}
which identifies the amplitude matching the experimental state.

In order to compare the catability with the well-established fidelity, we introduce the normalized infidelity of the experimental state~$\hat{\rho}$ with the ideal target state~$\ket{C_{\theta} (\alpha)}$ as
\begin{equation}\label{eq:infidelity_metric}
    \zeta (\alpha) \equiv 
    \frac{
        1 - \braket{C_{\theta} (\alpha) | \hat{\rho} | C_{\theta} (\alpha)}
    }{
        \min\limits_{\ket{\psi} \in \mathscr{H}} (
            1 - \abs{\braket{C_{\theta} (\alpha) | \psi}}^{2}
        )
    },
\end{equation}
which is normalized with respect to the class of Gaussian states~$\mathscr{H}$. Its value depends on prior knowledge of~$\alpha$. This can be, once again, resolved by using the global infidelity
\begin{equation}\label{eq:global_infidelity}
    \zeta = \min_\alpha \zeta(\alpha)
\end{equation}
which selects the best-matching amplitude for the state.

\begin{figure}[h]
    \includegraphics[width = \columnwidth]{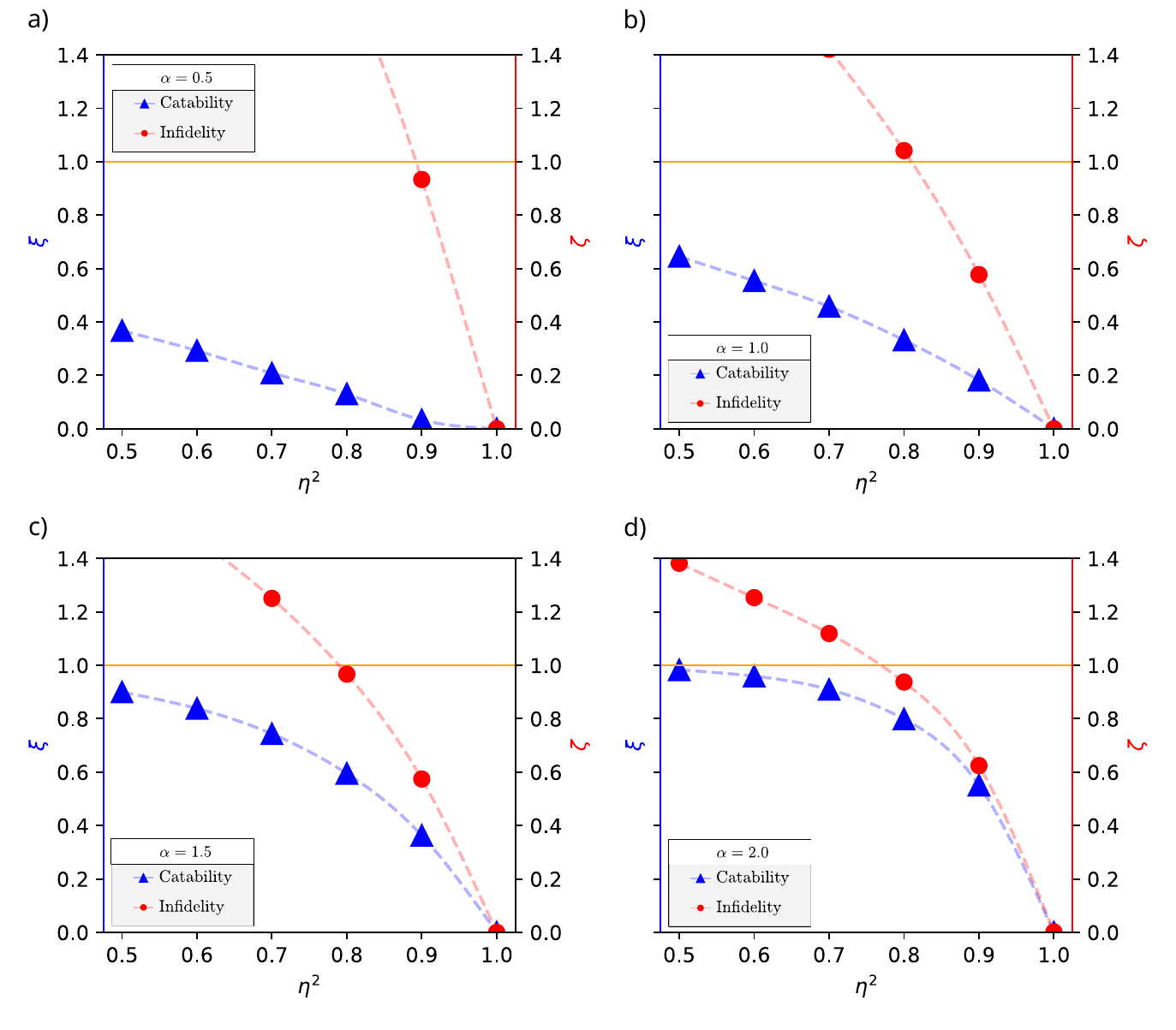} 
    \caption{
        Resilience global generalized catability~$\xi$ (blue) and fidelity~$\zeta$ (red) against loss, expressed in terms of intensity transmittance~$\eta^{2}$. Each panel depicts a different amplitude~${\alpha \in \{ 0.5, 1.0, 1.5, 2.0 \}}$. The presented results were obtained numerically on a truncated Fock basis with dimension~${N = 70}$.
    }
    \label{fig:catability_infidelity}
\end{figure}

The two quantifiers, $\xi$ and $\zeta$, offer different degrees of resilience against loss degrading the experimental states. We investigate their behavior by considering ideal states decohered by pure loss~\cite{lejeannic2018}, which can be modeled using an imbalanced beam splitter with intensity transmittance ${\eta^{2} \leq 1}$. The channel has an equivalent Kraus operator decomposition~\cite{ivan2011}
\begin{equation}
    \hat{M}_{k} (\eta) = 
        \sqrt{
            \frac{(1 - \eta^{2})^{k} }{k!} 
        } 
        \eta^{\hat{n}} 
        \hat{a}^{k}
        ,
\end{equation}
which can applied to the initial Kerr cat state $\ket{C_{\theta}(\alpha)}$ as 
\begin{equation}
    \hat{\rho} ({\eta}) = 
    \sum_{n=0}^{\infty} 
        \hat{M}_{k} (\eta)
        \ket{C_{\theta}(\alpha)} \bra{C_{\theta}(\alpha)}
        \hat{M}_{k}^{\dagger} (\eta)
    .
\end{equation}
The transformed states are then evaluated using the global quantifiers, $\xi$ and $\zeta$. 

We consider the states $\ket{C_{\pihalf}(\alpha)}$ and present the quantifiers as functions of transmittance~$\eta^{2}$ in~\figref{fig:catability_infidelity} for states a number of amplitudes~$\alpha$. Following the analysis presented within~\figref{fig:expecation_norm} we only consider displacing their positive fringes, hence we use the parameters~${s = 1}$ and~${\beta \geq 0}$ in~\eqref{eq:catability_metric}. The states~$\ket{C_{- \pihalf}(\alpha)}$ produce identical values when~${\beta \leq 0}$.
The presented results reveal that the overall catability exhibits greater resilience against loss, especially so for smaller amplitudes where it outperforms infidelity by a long stretch.

\section{Practical catability}
\label{sec:practical}

A certainly practical advantage of the generalized catability~\eqref{eq:catability_metric} is that is does not, in principle, require a full tomographical reconstruction of the analyzed state. The expectation value of the operator~\eqref{eq:catability_operator} can be determined from a finite set of photon-number measurements. 

A particular decomposition of the operator~\eqref{eq:catability_operator} is given by
\begin{equation}\label{eq:catability_of_DnD}
    \begin{split}
        \hat{O}_{N}  
        &= 2 \hat{n}^2  + \abs{ \alpha} ^2 (4 \hat{n} + 1)  + 2\abs{\alpha}^4 - \hat{n} \\ 
        &-\frac{1}{2} \big( \hat{D}^{\dagger} (\alpha) \hat{n}^2 \hat{D}(\alpha) + \hat{D}^{\dagger} (-\alpha) \hat{n}^2 \hat{D}(-\alpha)  \big ) \\
        &+ \gamma \big( N^{(\beta)} -s \hat{D}(\beta) \hat{\Pi} \hat{D}^{\dagger}(\beta) \big).
    \end{split}
\end{equation}
The measurement statistics of the four constituent operators, 
\begin{gather}
    \hat{n},\; \hat{D}^{\dagger} (\alpha) \hat{n}^2 \hat{D}(\alpha),\;
    \hat{D}^{\dagger} (-\alpha) \hat{n}^2 \hat{D}(-\alpha), \\
    \text{and}\;\hat{D}^{\dagger} (-\beta) (-1)^{\hat{n}} \hat{D}(-\beta),
\end{gather}
are merely functions of the probabilities 
\begin{equation}
    p_{n} (\mu)
    = \text{Tr}[
        \hat{D}^{\dagger} (\mu) 
        \ket{n}\!\bra{n} 
        \hat{D}(\mu) 
        \hat{\rho}
    ]
\end{equation} 
for any quantum state $\hat{\rho}$ and can be used to determine the expectation value of the operator~\eqref{eq:catability_of_DnD} as
\begin{equation}
\begin{split}
    \langle \hat{O}_{N} \rangle
    &= \sum_n p_n(0) \big(2 n^2 - (1 - 4 \abs{\alpha}^2)n \big) \\
    & -\frac{1}{2} \bigg(\sum_n p_n(\alpha) n^2 + \sum_n p_n(-\alpha) n^2 \bigg ) \\
    & - \gamma  \sum_n p_n( -\beta) (-1)^n \\
    & + 2 \abs{\alpha}^4 + \abs{\alpha}^2 + \gamma N^{(\beta)}. 
\end{split}
\end{equation}
The only difference from the original catability~\cite{catability} is the inclusion of~$\beta$, which is a function of $\alpha$ and requires an additional measurement of the $\beta$-displaced photon-number operator.

\FloatBarrier
\section{Conclusion}\label{sec:conc}

We extended the concept of catability~\cite{catability} for conventional cat states to include the family of parity-less Kerr cats by modifying the original cat state nullifier to reflect the essential features of the parity-less states. Instead of relying on the expectation value of a plain parity operator, equal to sampling the Wigner function at the center of the phase space, we employ a suitably displaced parity operator, which allows us to approximate their fundamental interference structure.

Our approach preserves the spirit of the original catability. The modified nullifier can be used to define a witness of non-Gaussianity, based on features of the parity-less cat states, to find the best approximate Kerr states on finite dimensions, and to determine the quality of experimental Kerr cat states. From a practical standpoint, the nullifier can be evaluated from a finite set of photon-number measurements and does not require full tomography of the analyzed state. 

We compared its resilience against loss with fidelity, a standard tool used in experiments, and concluded that catability generally performs better, especially for states with lower amplitude.

We also identified a non-physical nullifier that captures the symmetry of the parity-less Kerr cat states exactly by employing an anti-linear operator of complex conjugation instead of the displaced parity operator. Even though this nullifier cannot be measured directly, it can be computed numerically for states that were fully reconstructed and serve as an ideal reference.

\begin{acknowledgments}
    The authors acknowledge use of the computational cluster of the Department of Optics. Several open-source software libraries~\cite{hunter2007,lam2015,harris2020,virtanen2020} were used in the numerical computations presented within the manuscript. 

    \v{S}B acknowledges the support from the Carlsberg Semper Ardens project QCooL. PM, JP, and MM acknowledge the financial support of the Czech Science Foundation (project 25-17472S), and European Union’s HORIZON Research and Innovation Actions under Grant Agreement no. 101080173 (CLUSTEC). MM acknowledges IGA-PrF-2026-005. JP and MM acknowledge the Quantera project CLUSSTAR (8C24003) of MEYS, Czech Republic. Project CLUSSTAR has received funding from the European Union’s Horizon 2020 Research and Innovation Programme under Grant Agreements No. 731473 and No. 101017733 (QuantERA). PM acknowledges a grant from the Programme Johannes Amos Comenius under the Ministry of Education, Youth and Sports of the Czech Republic reg. no. CZ.02.01.01/00/22\_008/0004649.
\end{acknowledgments}

\section*{Data availability}
All the supporting datasets will be published in a dedicated repository.

\appendix

\section{Exact nullifier of Kerr cat states}
\label{sec:app_a}

The conventional cat states put their well-defined parity on display. The support of even cat states includes only even Fock states. Conversely, odd cat states are composed of odd Fock states only. Their distinctive anatomy can be exploited for their identification using a parity operator.

Even though the parity-less Kerr cat states do not exhibit such complementary structure, their components are complex conjugate with each other. The states can still be distinguished from each other with an anti-linear operator of complex conjugation~\cite{wigner,cejnar2015,sakurai,ballentine}, which requires careful handling. 

The most distinctive behavior between linear and anti-linear operators is in their effect on states~${\ket{\psi} \in \mathscr{H}}$ and their dual linear maps~${\bra{\psi} \in \mathscr{L}(\mathscr{H})}$. Because~$\bra{\psi}\hat{A}$ is anti-linear, it is necessary to introduce complex conjugation 
\begin{equation}
    (\bra{\phi} \hat{A}) \ket{\phi} 
    = [\bra{\phi}(\hat{A} \ket{\psi})]^{*}
\end{equation}
when changing the direction in which the operator acts.

We can define the anti-linear operator~$\hat{K}$ of complex conjugation by its action on states~${\ket{\psi} = \sum_{i} c_{i} \ket{i} \in \mathscr{H}}$ with
\begin{equation}\label{eq:K}
    \hat{K} \ket{\psi} 
    = \hat{K} \sum_i c_i \ket{i} 
    = \sum_i c_i^* \ket{i}.
\end{equation}
It anti-commutes with the complex unit ${\hat{K} \imath \hat{K}^{-1} = - \imath}$, and consequently, with all complex numbers. Its action on coherent states, ${\hat{K} \ket{\alpha} = \ket{\alpha^*}}$, follows directly from its definition~\eqref{eq:K}.

The operator~$\hat{K}$ can be used to define the exact nullifier only for certain amplitudes~$\alpha$, specifically when either~${\re{\alpha} = 0}$ or~${\im{\alpha} = 0}$. The modified $\hat{O}_{P}$ operators then become
\begin{alignat}{3}
    \label{eq:im_alpha}
    \hat{O}^{\text{Im}{(\alpha)}}_{P} (\gamma,s) 
    & \equiv \gamma(\imath -s \hat{K}) 
    & \quad\text{for}\; \re{\alpha} = 0,\\
    \label{eq:re_alpha}
    \hat{O}^{\text{Re}{(\alpha)}}_{P} (\gamma,s) 
    & \equiv \gamma(\imath -s \hat{\Pi}\hat{K}) 
    & \quad\text{for}\; \im{\alpha} = 0,
\end{alignat}
where ${s = \pm 1}$ denotes the sign in~\eqref{eq:general_cat}.

In order to prove that~\eqref{eq:re_alpha} selects the correct subspace of ground states it is sufficient to show that
\begin{equation}\label{eq:A1}
    \braket{
          C_{\pm \pihalf}(\alpha )
        | (\imath - s \hat{K}) 
        | C_{\pm \pihalf}(\alpha )
    } = 0
\end{equation}
selects the c subspace of the parity-less Kerr cat states $\ket{C_{\pm\pihalf}(\alpha )}$ for $\alpha \in \mathbb{C}$ and $\re{\alpha} = 0$.
Using the anti-commutation property of $\hat{K}$ with a complex unit and the action on coherent states, we have
\begin{equation}
    \begin{aligned}
        f(\theta) 
        & =
        \braket{
             C_{\theta}(\alpha )
             | \hat{K}
             |  C_{\theta}(\alpha ) 
        } \\
        & = \frac{ 
            \emath^{- \abs{\alpha}^{2} }
        }{M} \left(
            \emath^{\alpha^{*2} } 
            + \emath^{-2 \imath \theta + \alpha^{*2} } 
            + 2 \emath^{-\imath  \theta - \alpha^{*2} }
        \right)
    \end{aligned}
\end{equation}
where $M$ denotes the normalization~\eqref{eq:norm}. We obtain
\begin{equation}
    f\left(\pm\frac{\pi}{2}\right) = \mp \imath \emath ^{-\abs{\alpha}^2 - \alpha^{*2} }
\end{equation}
by using the relation for the amplitude probability of the two different coherent states $\ket{\alpha}$ and $\ket{\alpha '}$ \cite{leonhardt}
\begin{equation}
    \braket{\alpha'|\alpha} = 
    \emath^{
        -\frac{\abs{\alpha}^2}{2} -\frac{\abs{\alpha'}^2}{2} + \alpha'^* \alpha 
    }.
\end{equation}
When ${\alpha \in \mathbb{C}}$ and ${\re{\alpha} = 0}$ we obtain
\begin{equation}
    \braket{
        C_{\pm \pihalf}(\alpha )
        | \hat{K}
        | C_{\pm \pihalf}(\alpha )
    }
    = \mp \imath.
\end{equation}
The equation \eqref{eq:A1} is therefore valid for $s = \pm 1$. \QED

Similar steps can be taken to show that~\eqref{eq:im_alpha} satisfies
\begin{equation}
\braket{
        C_{\pm \pihalf}(\alpha )
        | (\imath - s \hat{\Pi}\hat{K})
        | C_{\pm \pihalf}(\alpha )
    }
    = 0
\end{equation}
for $\alpha \in \mathbb{C}$ with $\im{\alpha} = 0$.

\vfill
\clearpage
\bibliography{biblio}
\end{document}